\documentclass[acmsmall,nonacm]{acmart}

\AtBeginDocument{%
  }

\setcopyright{none}
\copyrightyear{}
\acmYear{}
\acmDOI{}

\begin{document}

\title{Deontic Paradoxes in Library Lending Regulations: A Case Study in Flint}
\author{Sterre Lutz}
\email{s.lutz@students.uu.nl}
\orcid{}
\affiliation{%
  \institution{Utrecht University}
  \streetaddress{Postbus 80125}
  \postcode{3508 TC}
  \city{Utrecht}
  \country{Netherlands}
}
\affiliation{%
    \institution{TNO}
    \streetaddress{Postbus 23}
    \postcode{3769 ZG}
    \city{Soesterberg}
    \country{Netherlands}
}

\begin{abstract}
Flint is a frame-based and action-centered language developed by \citet{van2016calculemus} to capture and compare different interpretations of sources of norms (e.g. laws or regulations). The aim of this research is to investigate whether Flint is susceptible to paradoxes that are known to occur in normative systems. The example of library lending regulations -- first introduced by \citet{sergot1982prospects} to argue for including deontic concepts in legal knowledge representation -- is central to this analysis. The hypothesis is that Flint is capable of expressing \citeauthor{sergot1982prospects}'s library example without the occurrence of deontic paradoxes (most notably: the Chisholm paradox \cite{ronnedal2019contrary}). This research is a first step towards a formal analysis of the expressive power of Flint as a language and furthers understanding of the relation between Flint and existing deontic logics. 
\end{abstract}

\maketitle
\section{Introduction}
Sources of norms, such as legal texts, are generally described in natural (albeit sometimes domain-specific) language. Natural language often leaves room for different `readings' of the same text, leading to more than one interpretation of the same source. For example, \citet{allen1994hohfeld} demonstrated that their Interpretation-Assistance program could generate 2560 structurally different interpretations of a normative text of only four lines of natural language. Additionally, to get a complete overview of the norms in a domain, one might want to consult multiple sources, and different interpretations can also come forth based on the selection and relative valuation of those possibly conflicting sources. 

Flint is a frame-based language geared towards capturing specific interpretations of sources of norms. It was developed by \citet{van2016calculemus} to facilitate structured discussion and comparison of different interpretations. Potential use cases can be found far and wide: for example, the legislative branch of an organization when creating or evaluating normative texts, as well as the executive branch responsible for complying with the texts (e.g. the Dutch Aliens Act \cite{doesburg2019explicit}). The normative sources can vary: most prominently analyzed are legal texts, but Flint can also be applied to sources like the semi-formal regulations of a library loaning books to members, or even the verbal agreements made when starting a book club with friends.


The aim of this research is to investigate whether Flint is susceptible to paradoxes that are known to occur in normative systems (most notably: the Chisholm paradox \cite{ronnedal2019contrary}). The method is creating a Flint model of library lending regulations -- which have been shown to contain the Chisholm paradox when expressed in Standard Deontic Logic \citep{jones1990deontic}\citep{herrestad1991norms} -- and determining whether the paradoxes occur in that model. This research is a first step towards a formal analysis of the expressive power of Flint as a language and furthers understanding of the relation between Flint and existing deontic logics. 



\section{Flint}
A Flint frame consists of several fixed components, each to be assigned a value to create an instance of a frame. The language is limited to three classes of frames: acts, facts and duties. An act frame describes what some actor can do and what happens because of that action, given that some preconditions are satisfied. If these preconditions are complex (e.g. `A OR B'), a fact frame can be used to explicitly give the (Boolean) formula for this complex fact and support this definition with one or more sources. A duty frame describes some desirable or ideal course of action, by enumerating three acts: the act that creates the duty (e.g. borrowing a book), the act that enforces the duty (e.g. giving a fine), and the act that terminates the duty (e.g. returning the book). 

In Flint, duties can only concern acts. It is not possible to express which facts should be true or false. Thus, it is an ought-to-do language, not an ought-to-be language like Standard Deontic Logic. Another important characteristic of Flint, is that duties and facts are created (made true) or terminated (made false) by an act. The acts therefore change the truth-assignments of facts and duties, similar to Propositional Dynamic Logic (PDL) and its deontic extension PDeL \citep{meyer1988different}. All in all, Flint clearly distinguishes acts and facts, and has a distinctly dynamic nature, which will play an important role in the preliminary analysis. 

\section{Deontic Paradoxes}

\citet{sergot1982prospects} posed that even for formalizing a very limited set of library regulations -- a selection from the Library Regulations at Imperial College -- some notion of obligation was required. Consequently, both \citet{jones1990deontic} and \citet{herrestad1991norms} showed that even that limited set of regulations could result in the Chisholm paradox when expressed in SDL, due to the following set of rules:
\begin{enumerate}
    \item X SHALL RETURN Y BY DATE DUE.
    \item IF X RETURNS Y BY DATE DUE THEN DISCIPLINARY ACTION SHALL NOT BE TAKEN AGAINST X.
    \item IF X DOES NOT RETURN Y BY DATE DUE THEN DISCIPLINARY ACTION SHALL BE TAKEN AGAINST X.
\end{enumerate}
When these rules are faced with a person who does not return their book, a contradiction occurs in SDL. Jones posed that a more `sophisticated' extension of deontic logic is required to prevent this contradiction. Such extensions (e.g. the addition of temporal modalities \citep{brunel2006state} or agency \cite{meyer1988different}) circumvent paradoxes such as the Chisholm paradox with varying degrees of success \citep{ronnedal2019contrary}.

\section{Preliminary Analysis and Hypothesis}
Previous case studies (e.g. \citep{doesburg2019explicit}) show that Flint is capable of describing regulations similar to those presented by \citet{sergot1982prospects}. Moreover, the distinction in Flint between acts and facts, and the dynamic nature of facts and duties (they are not \textit{implied} by actions, but \textit{created}) seems to largely align with PDeL. For this deontic version of dynamic logic, \citet{meyer1988different} has shown that the type of expressions that can lead to the Chisholm paradox (and similar paradoxes) can be expressed without contradictions.


Based on the preliminary analysis, the hypothesis is that Flint is capable of expressing \citeauthor{sergot1982prospects}'s library example without the occurrence of the Chisholm paradox. 
\bibliographystyle{ACM-Reference-Format}
\bibliography{sources}

\appendix


\end{document}